\documentclass[letterpaper, 10 pt, conference]{ieeeconf}

\IEEEoverridecommandlockouts

\usepackage[utf8]{inputenc}
\usepackage{textcomp}
\usepackage{microtype}
\usepackage{booktabs} %
\usepackage{hyperref}

\usepackage{xcolor}
\usepackage{fancybox}
\usepackage{graphicx}
\usepackage{amsmath}
\usepackage{amsfonts}
\usepackage{amssymb}
\usepackage{gensymb}
\usepackage{mathtools}
\usepackage{eurosym, booktabs}
\usepackage{pgfgantt}
\usepackage{colortbl}
\usepackage{caption}
\usepackage{subcaption}
\usepackage{float}
\usepackage{algorithmicx}
\usepackage{algorithm}
\usepackage{algpseudocode}
\usepackage{siunitx}
\usepackage{bbold}
\usepackage{nicefrac}

\usepackage{amsthm}
\usepackage{mathptmx}
\usepackage{etoolbox}
\usepackage{tabularx}
\usepackage{multirow}

\theoremstyle{definition}
\newtheorem{definition}{Definition}[section]

\makeatletter
\newcommand{\linebreakand}{%
\end{@IEEEauthorhalign}
\hfill\mbox{}\par
\mbox{}\hfill\begin{@IEEEauthorhalign}
}
\makeatother

\makeatletter
\newcommand\fs@spaceruled{\def\@fs@cfont{\bfseries}\let\@fs@capt\floatc@ruled
  \def\@fs@pre{\vspace{\baselineskip}\hrule height.8pt depth0pt \kern2pt}%
  \def\@fs@post{\kern2pt\hrule\relax}%
  \def\@fs@mid{\kern2pt\hrule\kern2pt}%
  \let\@fs@iftopcapt\iftrue}
\makeatother

\usepackage{cite} %

\usepackage[nolist]{acronym}
\begin{acronym}
  \acro{NN}{neural network}
  \acro{DNN}{deep neural network}
  \acro{RNN}{recurrent neural network}
  \acro{ML}{machine learning}
  \acro{PIML}{physics-informed machine learning}
  \acro{MPC}{model predictive control}
  \acro{RL}{reinforcement learning}
  \acro{GP}{Gaussian process}
  \acroplural{GP}[GPs]{Gaussian processes}
  \acro{MILP}{mixed-integer linear programming}
  \acro{BO}{Bayesian optimization}
\end{acronym}

\title{\textbf{Physics-Informed Machine Learning\\
for Modeling and Control of Dynamical Systems}}

\author{Truong X. Nghiem$^{1}$, J\'an Drgo\v na$^{2}$, Colin Jones$^{3}$, Zoltan Nagy$^{4}$, Roland Schwan$^{3}$, Biswadip Dey$^{5}$,\\
Ankush Chakrabarty$^{6}$, Stefano Di Cairano$^{6}$, Joel A. Paulson$^{7}$, Andrea Carron$^{8}$, Melanie N. Zeilinger$^{8}$\\
Wenceslao Shaw Cortez$^{2}$, and  Draguna L. Vrabie$^{2}$
    \thanks{$^{1}$School of Informatics, Computing, and Cyber Systems, Northern Arizona University, Flagstaff, AZ 86011, USA {\tt\small truong.nghiem@nau.edu}}%
    \thanks{$^{2}$Pacific Northwest National Laboratory, Richland, WA, USA {\tt\small \{jan.drgona, w.shawcortez, draguna.vrabie\}@pnnl.gov}}%
    \thanks{$^{3}$EPFL, Switzerland {\tt\small \{colin.jones, roland.schwan\}@epfl.ch}}%
    \thanks{$^{4}$The University of Texas at Austin, USA {\tt\small nagy@utexas.edu}}%
    \thanks{$^{5}$Siemens Corporation, Technology, Princeton, NJ 08540, USA {\tt\small biswa-dey@ieee.org}}%
    \thanks{$^{6}$Mitsubishi Electric Research Laboratories, Cambridge, MA 02139, USA. {\tt\small achakrabarty@ieee.org, dicairano@ieee.org}}%
    \thanks{$^{7}$The Ohio State University, Columbus, OH 43210, USA {\tt\small paulson.82@osu.edu}}%
    \thanks{$^{8}$ETH Zurich, Switzerland {\tt\small {carrona,mzeilinger}@ethz.ch}}%
    \thanks{This material is based upon work supported by the National Science Foundation under Grant No.~2138388 and Grant No.~2238296.}
}

\begin{document}
\maketitle

\begin{abstract}

Physics-informed machine learning (PIML) is a set of methods and tools that systematically integrate machine learning (ML) algorithms with physical constraints and abstract mathematical models developed in scientific and engineering domains. As opposed to purely data-driven methods, {PIML} models can be trained from additional information obtained by enforcing physical laws such as energy and mass conservation. More broadly, {PIML} models can include abstract properties and conditions such as stability, convexity, or invariance. 
The basic premise of {PIML} is that the integration of ML and physics can yield more effective, physically consistent, and data-efficient models. 
This paper aims to provide a tutorial-like overview of the recent advances in {PIML} for dynamical system modeling and control. 
Specifically, the paper covers an overview of the theory, fundamental concepts and methods, tools, and applications on topics of: 1) physics-informed learning for system identification; 2) physics-informed learning for control; 3) analysis and verification of {PIML} models;  and 4) physics-informed digital twins.  
The paper is concluded with a perspective on open challenges and future research opportunities. 
\end{abstract}

\section{Introduction}

\newcommand{\sdcComment}[1]{ {\bf \color{blue}(SDC: 
#1)\color{black} }}

\newcommand{\sdcHide}[1]{ {\bf \color{gray}(SDC: 
#1)\color{black} }}

Modern engineering systems generate large amounts of data either via sensors in the physical world or via simulation of virtual environments.
The increased storage and computational power of the underlying hardware and communication infrastructure paved the way for the use of data-driven algorithms.
 Machine learning (ML) methods leverage a large amount of data to achieve remarkable success, especially 
in  areas such as games, speech 
recognition, or image processing. 
These recorded successes especially occur in the areas where there is an abundance of data and where the
underlying processes 
have hard-to-discover governing laws and are driven 
by non-obvious fundamental principles. In such cases, ML shows 
strong capabilities in learning non-obvious relations that allow 
to achieve the desired tasks if a sufficient amount of representative data is available.

Due to such successes, a number of 
investigations have started involving the application of ML to other domains, including physical systems, such as 
automotive~\cite{Balaji2020}, aerospace~\cite{Brunton_aerospace_2021}, 
process control~\cite{NIAN2020106886}, energy systems~\cite{ML_energy_2021}, and robotics~\cite{OpenAI_rubik_cube}. 
However, these traditional engineering applications 
are governed by fundamental principles and physical constraints 
that have been studied for centuries and are thus better understood or well-known. Some examples 
include  conservation laws of mass or energy, 
fundamental laws of motion or electromagnetism, ranges of 
constants such as efficiencies and physical dimensions or gravitational constants.
For physical systems, there may be expectations that ML models
will satisfy such principles and constraints.
 While the prediction accuracy of ML methods may be as 
good in the computer science domains, the lack of satisfaction of 
fundamental principles and underlying physics and safety constraints represent a significant limitation 
to actual deployment of ML models in real-world applications. 

Properly enforcing  physical laws and constraints in ML 
may result in several benefits, affecting training, performance, 
and trust of the ML result such as: $(i)$ 
reduced data requirement of the ML model due to learning in lower dimensional, 
manifolds; $(ii)$ higher precision and improved generalization since the underlying physical principles will be  satisfied and the errors 
related to violating them will be avoided; $(iii)$ increased interpretability and trust by the application engineers by upholding the known domain principles.
However, most of the standard ML approaches cannot leverage physical principles and constraints, resulting in a limited real-world impact of ML in the safety-critical domains~\cite{RL_challenges2019}. 

The cautious adoption of modern ML methods in real-world engineering domains is a direct result of the research priorities in the historical developments within the computer science domains.
More specifically, ML methods have been primarily 
developed in domains related to human aspects and behaviors,
such as vision or language recognition,
where the safety requirements are less strict. Furthermore, those domains consist of
underlying physical principles that are still partially unexplained or extremely complex to describe and hence hard to be embedded directly in the ML algorithms.

In recent years, researchers from both computer science and engineering domains have recognized the potential of bridging this gap by developing new methods that combine algorithms and tools from machine learning with well-known engineering models and principles.
This quest to extend machine learning to account for the physical 
principles and constraints of the underlying process gave rise to the 
so-called Physics Informed Machine Learning (PIML), also referred to as Scientific Machine Learning (SciML).
Related surveys in this area include PIML methods for  partial differential equations (PDEs)~\cite{karniadakisPhysicsinformedMachineLearning2021} or generic dynamical systems~\cite{Wang2021,Legaard2023}.
However, to the author's best knowledge, there is no tutorial overview of the PIML methods for the control of dynamical systems.
This paper aims to bridge this gap.
The rest of the paper is structured as follows, section II. reviews the landscape of PIML methods for modeling and control of dynamical systems, including PIML methods for system identification, control design, and formal verification. 
Section III. briefly summarizes current challenges and opportunities. While section IV. concludes the paper.

\section{Landscape of Physics-Informed Machine Learning Methods for Modeling and Control}

In this paper, we define {PIML} as follows.
\begin{definition}
{{PIML} is a set of methods and tools that systematically integrate recent advancements in machine learning algorithms with mathematical models developed in various scientific and engineering domains.
As opposed to pure data-driven methods that assume no existence of prior domain knowledge, 
{PIML} models can be trained from additional information obtained by enforcing constraints such as symmetries, causal relationships, or conservation laws.  More broadly, {PIML} models can include abstract
properties and conditions such as stability, convexity, or invariance from domains such as dynamical systems and control theory.}
\end{definition}

\subsection{Physics-informed learning for system identification}
\label{sec:PIML-sysid}

One of the first applications of {PIML} in  control is to learn dynamical models for a physical system given some prior knowledge and time series data.
Conventional dynamical system and control theories usually use \emph{white-box } modeling to develop models from physical principles and system specifications, which often require extraordinary effort and expertise.
On the other hand, purely data-driven approaches employ \emph{black-box } modeling, which learns models purely from data without making any prior physical assumptions, relying on {ML} techniques such as \acp{NN} 
and regression trees (RT). 
While \emph{gray-box}  modeling and \ac{PIML} both aim to bridge the gap between the white-box and black-box modeling paradigms by integrating physical laws into models,
they take different approaches.
\emph{Gray-box} modeling starts from the same physical principles used in white-box modeling that are subsequently simplified  to obtain a reduced-order model structure, which can be used to fit data to identify its parameters.
Therefore the \emph{gray-box} approach still requires significant expert knowledge and often ignores nonlinear phenomena, causing accuracy loss.
On the other hand, \Ac{PIML} starts from the black-box modeling end and embeds prior knowledge of the system's physics into {ML} methods to yield models that are more interpretable, robust, accurate, and physically consistent for generalization tasks while maintaining the relative ease of use of \emph{black-box} {ML} methods.
This section reviews the main PIML approaches for incorporating physics into the data-driven modeling of dynamical systems. Namely, physics-informed model architecture,  
 physics-informed loss function design, or a combination of the two~\cite{karniadakisPhysicsinformedMachineLearning2021,Rackauckas2020GeneralizedPL}.

\begin{definition}
Model architecture represents an interconnected set of computational nodes that form the resulting computational graph defining the mathematical expression defining the ML model. Architectures of  ML models are often defined as composite functional forms or block diagrams. 
\end{definition}

\subsubsection{Structural priors in discrete-time models}
Discrete-time dynamical models are represented by a general class of state-space models (SSM) given as
\begin{equation} \label{eq:SSM}
       {x}_{k+1} = {f}_{\text{SSM}}\left({x}_k, {u}_k\right)
\end{equation}
with ${x}_k \in \mathbb{R}^{n_x}$ and
${u}_k \in \mathbb{R}^{n_u}$ being system states and control inputs, respectively. Most of the structural PIML approaches are concerned with designing the functional form  of~\eqref{eq:SSM} to offer increased expressivity compared to classical linear SSM while satisfying desired properties.

Recent examples include neural SSM~\cite{krishnan2016structured,NIPS2018_8004}, deep Koopman models~\cite{yeung2019learning,Lusch2018}, Hammerstein-Wiener neural models \cite{HW_RNN2008,OgunmoluGJG16}, graph neural network-based (GNN) time-stepper models~\cite{GNN_control2018,kipf2018neural}, 
 or SINDY-type methods based on sparse regression of candidate basis functions \cite{Brunton_2016,BRUNTON2016710}.
Further examples include architectures such as Non-Autonomous Input-Output Stable Nets (NAIS-Nets) with guaranteed asymptotic stability of its forward pass dynamics based on stable matrix factorization of the state dynamics, 
joint learning of the Koopman model representation  and state observer while ensuring observability~\cite{vijayshankar2022co}, or so-called Gumbel Graph Networks (GGN) for modeling network dynamics~\cite{Zhang2019}.
While authors in~\cite{DiNatale22b} introduced physically consistent \ac{NN} inspired by resistance-capacitance (RC) networks applied to modeling building thermal dynamics.

\subsubsection{Structural priors in continuous-time models}
Continuous-time models are, in general, represented by a set of ordinary differential equations (ODE) given as
\begin{equation} \label{eq:ode}
        \frac{d{x}}{dt} = {f}_{\text{ODE}}\left({x}, {u}\right)
\end{equation}
with $\mathbf{x} \in \mathbb{R}^{n_x}$ and
$\mathbf{u} \in \mathbb{R}^{n_u}$ being system states and control inputs, respectively.
Neural ODEs~\cite{Chen2018} have been proposed as black-box counterparts to white-box ODEs for data-driven modeling of dynamical systems. In NODEs, the right-hand side (RHS) of the differential equation~\eqref{eq:ode} is approximated by deep neural network $NN(\mathbf{x}, \theta)$ with trainable parameters $\theta$.
However, purely black-box NODEs may require prohibitively large training data and  may struggle with generalization and physical laws.
Researchers have been looking at ways to design the structure of the RHS in~\eqref{eq:ode} to balance conflicting criteria such as the expressivity of black-box models and the physical consistency of white-box models.

Examples include incorporated stability-promoting priors in NODEs via spectral element method~\cite{quaglino2020snode}, multiple-shooting integration schemes~\cite{Massaroli2021,Turan2022}, neural delay differential euquations~\cite{Schlaginhaufen2021}, 
or extensions supporting uncertainty quantification (UQ) such as Bayesian NODEs~\cite{BayesNODE2020}, stochastic NODEs~\cite{NeuralSDEs2021}, or jump stochastic NODEs capable of handling discrete events~\cite{Junteng_neuralJumpSDEs2019}.
Most recently, \textit{universal differential equations} (UDE)~\cite{Rackauckas2020} have been proposed as an extension to NODEs to allow for a systematic combination of white-box components and black-box components. Specific examples of PIML structured continuous models include the use of UDEs for modeling networked dynamical systems~\cite{Koch2023}, or continuous graph neural networks (CGNNs) \cite{CGNN2020} also called graph neural ordinary differential equations (GDEs) \cite{GDEs2019}.
Related approaches include \textit{energy-based} model architectures such as Hamiltonian~\cite{HamiltonianDNN2019,SymplecticODE2019} or Lagrangian~\cite{LagrancianDNN2019,Cranmer2020} neural networks, and their various extensions such as graph Hamiltonian neural networks (HNN)~\cite{sanchezgonzalez2019hamiltonian}, Hamiltonian dynamics with dissipative forces~\cite{zhong2019dissipative}, HNN with explicit constraints~\cite{Finzi2020}, or non-canonical Hamiltonian systems~\cite{eidnes2023pseudo}. Such approaches have also been extended to learn dynamics of rigid body systems with contacts and collisions~\cite{zhong2021extending} or to learn Lagrangian dynamics from high-dimensional video data~\cite{NEURIPS2020_79f56e5e}. 
Others have proposed jointly learning Neural Lyapunov functions together with the forward dynamics model~\cite{Manek2020}.
The main advantage of these 
 \textit{energy-based} approaches is that they can satisfy conservation laws or enforce properties such as stability or dissipativity by design.

  \subsubsection{Matrix factorizations} 
Linear algebra components are ubiquitous in   machine learning, including 
 weights in neural networks, operators such as Koopman or Perron-Frobenius, or Jacobian matrices of differential equations. 
 Many matrix factorization methods exist for designing desired properties such as sparsity, symmetry, positive definiteness, eigenvalue placement, or invertibility. 
Naturally, researchers have exploited  various matrix factorizations in weights of deep neural networks~\cite{DissipativeNets2022}, with 
  examples including Perron-Frobenius  \cite{tuor2020constrained}, orthogonal \cite{mhammedi2017efficient, jia2019orthogonal}, spectral \cite{zhang2018stabilizing}, symplectic \cite{haber2017stable,SymplecticODE2019}, anti-symmetric \cite{chang2019antisymmetricrnn}, Gershgorin disc \cite{lechner2020gershgorin}, and Schur Decomposition \cite{kerg2019non} weights.
Let's consider spectral factorization~\cite{zhang2018stabilizing} as an example. In this approach, the weight matrix $\mathbf{W} $ is parametrized by the components of
singular value decomposition (SVD) method as follows,
  \begin{subequations}
\begin{align}
    {{\boldsymbol\Sigma}} &= \text{diag}(\lambda_{\text{max}} - (\lambda_{\text{max}} - \lambda_{\text{min}}) \cdot \sigma(s))\\
   \mathbf{W} &= {\mathbf{U} {\boldsymbol\Sigma} \mathbf{V}} \label{eq:SVD}
\end{align}
\end{subequations}
Where ${\boldsymbol\Sigma}$ is a matrix of singular values, $\lambda_{\text{min}}$ and $\lambda_{\text{max}}$ 
represent lower and upper bounds for singular values, $\sigma$ is a sigmoid activation function,
$s$ is a vector of trainable singular value parameters, 
and $\mathbf{U}$ and $\mathbf{V}$ are trainable orthogonal matrices that can be either designed via Householder reflectors~\cite{zhang2018stabilizing} or via   
soft constraint penalties~\cite{Skomski2021}.
Authors in~\cite{DRGONA2021110992} used the matrix factorizations to enforce dissipativity in  neural SSM for modeling building thermal dynamics.

\subsubsection{PIML loss functions}
This approach incorporates physics through \emph{learning biases} by imposing appropriate penalties during learning~\cite{karniadakisPhysicsinformedMachineLearning2021}.
In its general form, the composite physics-informed loss function $ \mathcal{L}$ is defined as
\begin{equation}
    \mathcal{L}  = \sum_i^n \ell_{\text{data}}^i + \sum_j^m \ell_{\text{physics}}^j
\end{equation}
where $\ell_{\text{data}}^i$ represents 
 $n$ number of data-driven objective terms, $\ell_{\text{physics}}^j$ define  $m$ number of physics-based regularization terms.

Examples include using soft penalties on eigenvalues or singular values of the neural network weights \cite{erichsonPhysicsinformedAutoencodersLyapunovstable2019,Guanya2019} to promote stability, promoting boundedness, and smoothness~ \cite{Skomski2021}, using penalties to regularize black-box GNN components \cite{Sungyong2019}, regularizing error and stiffness estimates in NODEs for improved accuracy and speed~\cite{pmlr-v139-pal21a}, and penalizing the violations of the Lyapunov stability conditions as additional loss term in training NODEs \cite{LyaNet2022}.
Others have proposed  Jacobian regularization~\cite{finlay2020train} for promoting stability, or using
surrogate loss functions for faster training of NODEs
\cite{kelly2020learning}.

\subsubsection{Case Study: Incorporating physics in the model via structural priors}
\begin{figure}[]
        \centering
        \includegraphics[width=1.0\linewidth]{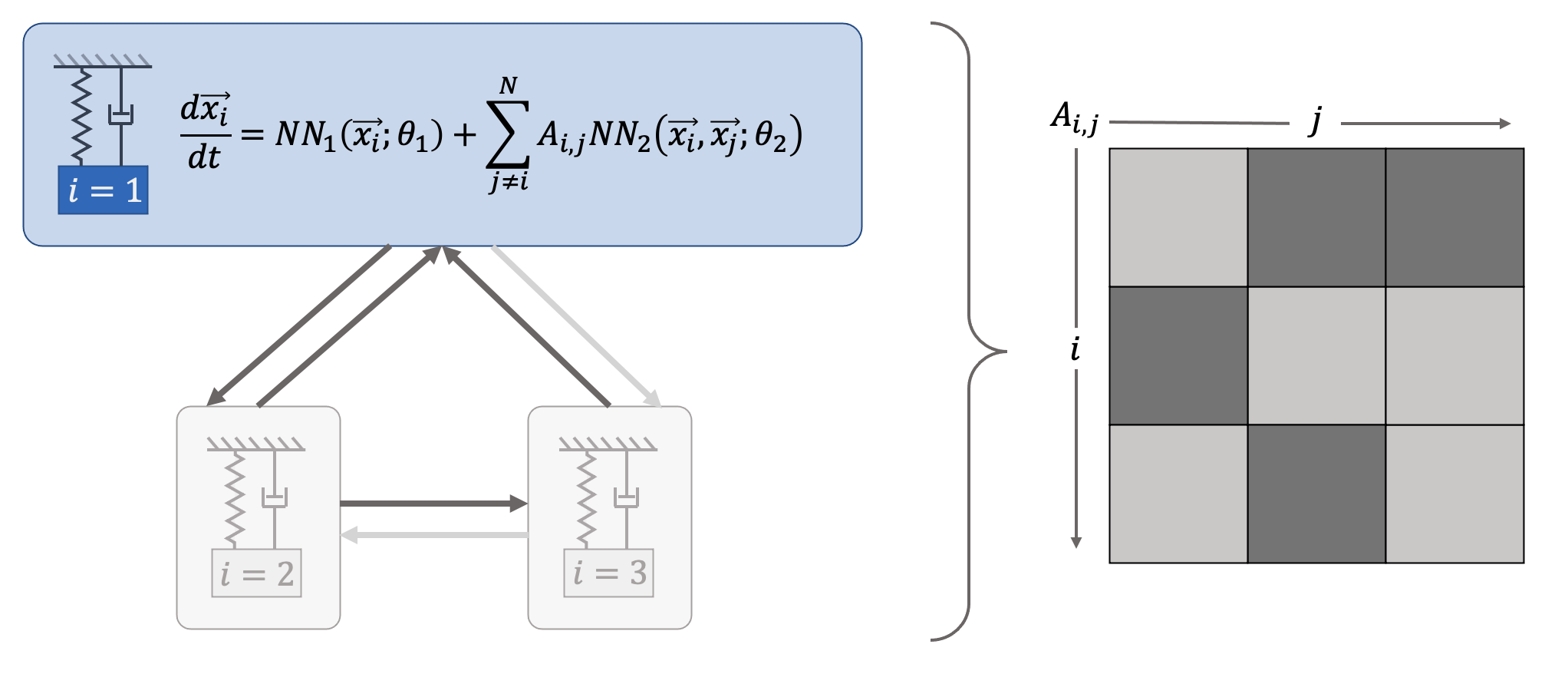}
	    \caption{Methodology of the  universal differential equation for learning components of networked dynamical systems~\cite{Koch2023}.}
		\label{fig:UDE_method}
\end{figure}

This case study is adopted from~\cite{Koch2023} 
and demonstrates the use of 
universal differential equations (UDE)~\cite{Rackauckas2020}
for data-driven modeling of networked dynamical  systems. The method illustrated in Fig.~\ref{fig:UDE_method}. 
Specifically, the UDE model used in~\cite{Koch2023} is given as follows,
\begin{equation} \label{eq:node}
        \frac{d{x}_i}{dt} = {NN_1}\left({x}_i;\theta_1\right) + \sum_{j\neq i}^N \mathbf{A}_{i,j}{NN_2}\left({x}_i, {x}_j;\theta_2\right) 
\end{equation}
The system dynamics is defined by the interaction of $N$ number of nodes represented by state vectors
${x}_i(t) \in \mathbb{R}^{n_x}$ where $i \in \mathbb{N}_1^{N}$ represents node index. Here ${NN_1}({x}_i,\theta_1): \mathbb{R}^{n_x} \rightarrow \mathbb{R}^{n_x}$ and ${NN_2}({x}_i,\theta_2): \mathbb{R}^{2n_x} \rightarrow \mathbb{R}^{n_x}$ are  neural networks modeling 
node and interaction dynamics, respectively. The trainable adjacency matrix $\mathbf{A} \in \mathbb{R}^{N \times N}$ defines the  connectivity between nodes, i.e.,  $A_{i,j} = 1$ means node $i$ is connected to node $j$. 
The equation \ref{eq:node} compactly represented as ${F}( {x}; \Theta)$ with lumped trainable parameters $ \Theta = \{\theta_1, \theta_2, \mathbf{A} \}$, can be solved with standard numerical ODE solvers,
\begin{equation} \label{eq:odesolve}
    {x}\left(t_{end}\right) = \texttt{ODESolve} \left( {F}\left( {x};\Theta \right), {x}_0, t_0, t_{end} \right) .
\end{equation}
In general, the gradients of ${F}( {x}; \Theta)$ can be computed in two ways via the adjoint sensitivity method as used in  NODE or via automatic differentiation of the computational graph of the unrolled ODE solver.
The loss function is formulated as regularized mean squared error (MSE) between predictions and measurements, given as,
\begin{equation} \label{eq:loss}
    \mathcal{L}  (\boldsymbol{\Theta}, \mathbf{A}) = \frac{1}{m} || \hat{\mathbf{X}}  - \mathbf{X}||^2_2 + \alpha|| \mathbf{A}||_1, 
\end{equation}
where ${\mathbf{X}}$ and $\hat{\mathbf{X}}$  are measured and predicted state trajectories, respectively. Scalar  $m$ represents a number of samples.
The second term represents $\ell_1$ penalty for promoting sparsity in the  adjacency matrix $\mathbf{A}$.

\begin{figure}[]
        \centering
        \includegraphics[width=1.0\linewidth]{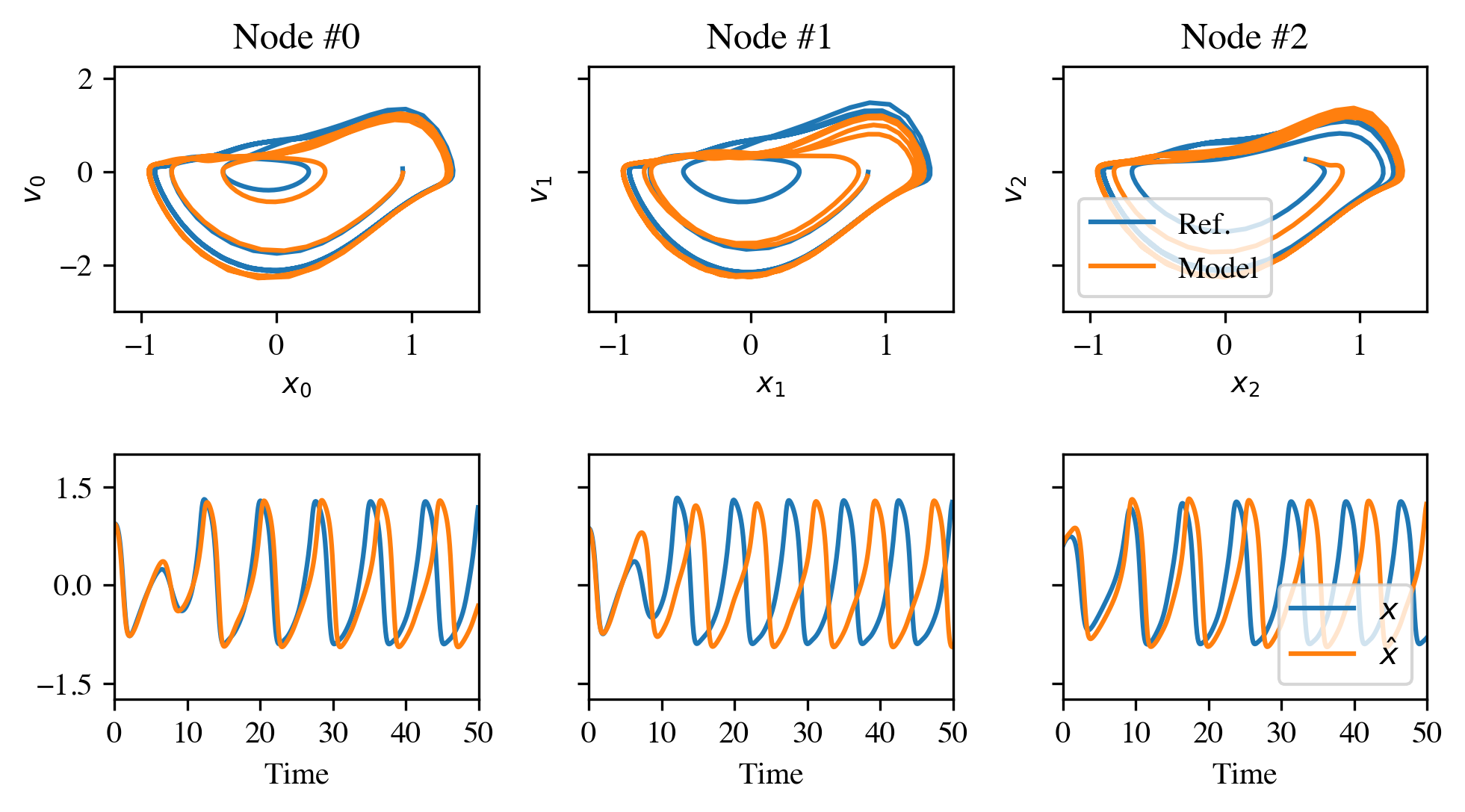}
	    \caption{Prediction performance of the UDE model~\ref{eq:node} (orange) on never-before-seen networked system (blue). }
		\label{fig:extrap}
\end{figure}

The authors in~\cite{Koch2023} demonstrate the utility of this UDE modeling approach in a case study with coupled nonlinear oscillators. The benefit of the UDE model in the form~\eqref{eq:node} is improved generalization and interoperability, as opposed to purely black-box NODE~\cite{Chen2018}. This is because the structure of~\eqref{eq:node} is closer to the governing physics of a general class of networked systems. 
To demonstrate their utility, the authors deploy the trained models on networked systems with never-before-seen topology. Fig.~\ref{fig:extrap} compares the phase portraits
and time series of the never-before-seen system against the predictions generated by the UDE model~\ref{eq:node} trained on data from a networked system with different topology. What is being demonstrated are the generalization capabilities of the UDE model~\ref{eq:node} to qualitatively reproduce physically plausible transient behavior, including the reconstruction of limit cycle attractors outside of the distribution of the training data. 

\subsection{Physics-informed learning for control}

\Ac{PIML} models have been widely used in model-based control methods, such as \ac{MPC} or model-based \ac{RL}.
This is typically done by combining physics-based priors with various regression methods to improve the control performance. %
Subsequently, with the spread of learning methods embedded with uncertainty quantification  (UQ) measures, robust and stochastic model-based controllers have been developed to exploit the uncertainty information provided by these \ac{PIML} models.
Including UQ methods makes the controller cautious and is of great significance where only a limited amount of data is available.
More recently, the use of \ac{PIML} has been extended to different areas such as learning control Lyapunov functions, tuning model-based controllers, designing dual control strategies, developing safe control frameworks, and learning explicit control policies. 
\ac{PIML} has also found applications in differentiable programming-based methods, which allows the designer to embed physical models and constraints while training a controller using automatic differentiation-based (AD) solvers, e.g., with physics-informed priors in actor-critics and model-based RL methods.

\subsubsection{\ac{PIML} dynamics models for \ac{MPC}}

Data-driven models have been widely used in \ac{MPC}; see \cite{Hewing2020a} for a comprehensive review of these methods up to 2020.
With recent advances, many approaches have been proposed for incorporating \ac{PIML} and physical priors into \ac{MPC}.
These methods employ learned \ac{PIML} dynamic models (Section~\ref{sec:PIML-sysid}) in the \ac{MPC} optimization formulation, therefore implicitly leveraging physical priors through the models.
Authors in~\cite{LenzKS15} used structured neural SSM in MPC to control the nonlinear behavior of a robotic hand in  cutting tasks. In~\cite{Nicodemus2021} a physics-informed \ac{NN}
is used with MPC in order to control a robotic arm.
Others have demonstrated the utility of using the sparse identification of nonlinear dynamics (SINDY) method for learning models for \ac{MPC}, in the so-called SINDY-MPC framework~\cite{Kaiser2008}. 
The use of ML models in \ac{MPC} with guarantees can be traced back to~\cite{ASWANI20131216, Bouffard2012}. The authors propose a robust control method that uses two models: the first is a simple physics-based linear model that accounts for the safety of the system  used to ensure constraint satisfaction, while the second is an ML model  used to maximize the performance of the controller. 
A different approach is presented in~\cite{Carron2019, Hewing2020b}, where a \ac{GP} is used to learn the error dynamics to improve the physics-based model and hence the controller performance. Another \ac{MPC} algorithm that uses kernel-based methods is presented in~\cite{Maddalena2021}. While the work in~\cite{Bunning2021} demonstrates the utility of even a simple physics-informed ARMAX model for building control tasks using MPC.

\subsubsection{Learning Lyapunov and Barrier functions} %
(Control) Lyapunov functions are crucial tools for determining the stability properties of dynamical systems, quantifying the domain of attraction and the robustness to perturbation, designing controllers, or designing terminal ingredients for \ac{MPC}. However, methods for synthesizing Lyapunov functions in closed-form for nonlinear dynamical systems, even with known models, are generally not available. 
Early approaches for learning Lyapunov functions with \acp{NN} date back to the 90s~\cite{long1993}. 
Recently, different authors~\cite{LyapunovNN2018,Manek2020} introduced neural Lyapunov functions, representing neural architectures that satisfy Lyapunov function properties by design.
Subsequently, authors in~\cite {LyapunovNN2018} have used these neural Lyapunov functions as verification tools for safe learning-based controllers, while~\cite{mittal2020neural} used  neural Lyapunov function for online tuning of MPC parameters, and in~\cite{LyapunovDPC22}  neural Lyapunov function was trained in conjunction with neural control policy. 
Another example is \cite{Grune2021}, where \acp{NN} are employed to learn compositional Lyapunov functions. This work takes inspiration from recent methods that solve high-dimensional PDEs using \acp{NN} by exploiting suitable structural properties. 
As Lyapunov functions can also be represented by PDEs, \ac{PIML} methods for learning PDEs can also be used to learn Lyapunov functions.
Control barrier functions (CBF) recently became a very popular computationally efficient method for safe learning-based control. However, similarly to Lyapunov functions, CBFs are in general, hard to design analytically for general nonlinear dynamical systems.
To alleviate this problem, different authors have proposed learning CBFs from data.
In~\cite{Kehan2021}, use \acp{NN} to approximate  the signed distance functions and subsequently use second-order cone programming to synthesize the control policy. 
Authors in~\cite{Robey2020} utilize imitation learning to learn CBFs from expert demonstrations of safe trajectories.
While others  present architectures for constructing neural barrier functions
\cite{qin2021learning} for multi-agent control, and 
robust neural Lyapunov-barrier~\cite{dawson22a} functions  for safe nonlinear control.

\subsubsection{Differentiable Control}
    These methods leverage automatic differentiation (AD)
    for computing gradients of the physics model or optimal control problem that can be used as part of the learning algorithm. 
    An example is to use AD for computing the backward gradients of  the underlying \ac{MPC} optimization problem. In principle, this can be done
    in two ways, by differentiating the KKT conditions constructed analytically~\cite{NEURIPS2018_ba6d843e} or by
    unrolling the computational graph of the \ac{MPC}  problem~\cite{OkadaRA17,Pereira2018}.
        The advantage of the methods based on differentiable programming (DP)~\cite{DiffProg2019} is that they allow for end-to-end training of different components of the optimal control problem. 
    Thanks to its versatility, this method has been used   
    to learn system dynamical models~\cite{East2020InfiniteHorizonDM}, weighting factors of the objective functions~\cite{NEURIPS2018_ba6d843e},  neural control policies~\cite{DRGONA202280,SDPC2022,Tabas2022}, or safety filters based on differentiable projections~\cite{Donti_control2020}, and differentiable control barrier function~\cite{Wei2022,Shuo2023}.
    Most recent applications of differentiable control include autonomous vehicles~\cite{karkus2022diffstack}, robotics~\cite{Xu-RSS-21,Degrave2016ADP}, building control~\cite{Gnu-RL2019,DRGONA202114}, traffic flow~\cite{Son2022differentiable}, epidemic processes~\cite{Asikis2022},
    or visuomotor control via differentiable rendering~\cite{gradsim}.
    Naturally, physical priors can be incorporated into the \ac{MPC} formulation through the system model, which will be taken into account when the optimization problem is differentiated.
    Most recently, several differentiable optimization and control libraries have emerged in the open-source domain~ \cite{cvxpylayers2019,gradu2020deluca,pineda2022theseus,Neuromancer2022}
   The DP approach has also been used to develop differentiable physics models to control partial differential equations (PDEs)~\cite{Kiwon2020,Holl2020Learning}.

\subsubsection{\Ac{PIML} for safe data-driven control}
Another use of \ac{PIML} models in combination with control-like algorithms is in safety frameworks. Given a safety-critical system, i.e., a dynamical system subject to state and input constraints, a safety framework %
can certify whether a proposed control input is safe to apply or not. Whenever a proposed input is unsafe, the safety framework can propose an alternative safe control input, which usually is as close as possible to the proposed one.
For instance, \cite{Wabersich2021,Tearle2021,9812259} present safety frameworks that use an \ac{MPC} like structure to determine the input that satisfies input and state constraints and is as close as possible to the proposed potentially unsafe input. 
Safety frameworks are also known under other names such as active set reachability~\cite{Gurriet2018}, Safety Handling Exploration with Risk Perception Algorithm (SHERPA)~\cite{Mannucci2018}, and model predictive safety filter~\cite{Wabersich2018}.
Other approaches use control barrier functions~\cite{Ames2019,barrier_DPC_2022} and reachability analysis~\cite{Chen2018,9146356} to ensure safety. A unified approach that brings predictive safety filters and control barrier functions, called predictive control barrier functions, has been recently presented~\cite{Wabersich2022a}

Safety frameworks have also been proposed for learning-based control, where \ac{ML} is used to learn either the system model or the control law.
\cite{Karg2020} uses an MILP formulation to train \ac{NN}-based controllers to satisfy input constraints, safety constraints, and stability conditions.
\cite{Donti2021a} uses differentiable projections to enforce Lyapunov stability conditions while minimizing a performance objective-based loss function (e.g., LQR) during training of a \ac{NN}-based control law.
Cautious control methods employ the uncertainty of the learned model to ensure safety constraints and become less conservative when the model is updated~\cite{Hewing2020b}.

\subsubsection{Physics-informed \ac{RL}}
\Ac{PIML} has been used in \ac{RL} to improve its accuracy and physical consistency.  Model-based reinforcement learning %
methods rely on a model of the environment and its dynamics with which the agent interacts. The models can be generated via high-fidelity physics-based simulations or learned from data. 
Deep-\ac{RL} methods employ \acp{NN} to generate data-driven models. Approaches similar to those presented in the \ac{MPC} section can be used to ensure physically consistent models.
For example, \cite{DiNatale22c} uses a physically consistent \ac{NN}~\cite{DiNatale22b} to learn a dynamical model of a building, which is then used with a deep \ac{RL} agent.
In \cite{Deisenroth2011}, a model-based policy search method, called PILCO, is developed where prior physics knowledge can be introduced in a \ac{GP} model.
Physics-informed \ac{RL} schemes informed by the underlying dynamics have been employed to tackle aircraft conflict resolution problems~\cite{PIRL_aircraft_2022} and power system optimization problems~\cite{PIRL_grid_2022,Wenqi2021}.
Recent works in~\cite{Gros2020,Zanon2021} provide a systematic framework for using \ac{RL} to tune the parameters of optimization-based \ac{MPC} policies. This RL-MPC framework combines the advantages of both approaches, namely the flexibility of \ac{RL} with constraints satisfaction and stability of \ac{MPC}. 
It has recently found applications in building energy system control~\cite{ARROYO2022118346}.

\subsubsection{Case study: Physics-informed safety filters for data-driven control}

The work~\cite{Wabersich2021} shows how to deploy RL algorithms on safety-critical systems, i.e., systems subject to state and input constraints, without violating constraints. The proposed approach makes use of predictive safety filters, namely, an optimization-based algorithm that receives the proposed control input and decides, based on the current system state, if it can be safely applied to the real system or if it has to be modified otherwise.
Given a dynamical system of the form~\eqref{eq:SSM}, the predictive safety filter solves each time-step the following optimization problem,
\begin{equation}
\label{pic:psf}
\begin{split}
    \min_{x,u} &\  \|u_{L} - u_{k}\|\\
    \text{s.t. } &x_{k+1} = f(x_{k}, u_{k}), \\
    & x_{k} \in \mathbb{X}, \ \ u_{k} \in \mathbb{U}, \ \
     x_{N} \in \mathcal{S}^t, \ \ x_{0} = x(t), 
\end{split}
\end{equation}
where $u_L$ is the input proposed by a potentially unsafe controller, e.g., an RL algorithm and $\mathcal{S}^t$ is a terminal invariant set for the dynamical system $f(x_k, u_k)$. The predictive safety filter minimizes the difference between the proposed input and a safe input in order to be as less invasive as possible. The input $u_{k}$ is then applied. 
While the aforementioned approach relies on a perfect knowledge of the system dynamics, in practice a perfect model is never available. The predictive safety filter can deal with the uncertain system by exploiting robust and stochastic model predictive control techniques. A parametric model 
$x_{k+1} = f(x_k, u_k, \theta)$ is considered
where $\theta$ is an unknown parameter vector that is either bounded or has a known prior distribution $p(\theta)$ with mean $\mathbb{E}[\theta]$.
In the aforementioned paper, the safety filter is applied to the swing-up of an inverted pendulum and the control of a quad-copter. In both cases, unsafe controllers are employed and the safety filter demonstrates how it is possible to satisfy constraints.

\subsection{Analysis and verification of \ac{ML} models}

In addition to system identification and control design, \ac{PIML} has been incorporated into the analysis and verification of dynamical systems.
Firstly, consider a learning structure, (e.g., \ac{NN}, \ac{GP}) on which \ac{PIML} methods typically rely on.
Properties of these structures, such as input-to-state stability, input-output bounds, and estimation of Lipschitz constants, are important to determine how these structures behave in a closed-loop setting.
Secondly, learned system dynamics would ideally have inherent physical properties, e.g., passivity, stability, and invariance, from the physical systems they are approximating.
Guarantees that such physical properties hold can be beneficial for future control design in addition to ensuring a better representation of the physical system itself.
Finally, learning-based control policies may lack guarantees of closed-loop stability or invariance.
Verification methods, including sampling-based and reachability, can be used to guarantee such desirable properties of the closed-loop system.
In the following, we survey the literature with respect to analysis and verification methods developed for \ac{PIML}.

\subsubsection{Analysis and verification of learned system dynamics}

These are methods for analyzing and verifying properties of learned models of dynamical systems in the open-loop setting.
They can be categorized by the class of system properties addressed as follows.
    
\paragraph{Stability}
 The stability and attractors of \acp{RNN} have been studied in continuous time in \cite{Zhang2014}.
Authors in~\cite{engelken2020lyapunov,vogt2020lyapunov} study Lyapunov spectra and Lyapunov exponents of \acp{RNN}.
Authors in~\cite{gler2019robust} empirically study stability of deep neural architectures in the context of solving forward-backward stochastic differential equations. %
\cite{DissipativeNets2022} analyzes the dissipativity of autonomous neural dynamics.
\cite{Abate2021} proposes a method for synthesis of Lyapunov \acp{NN} while providing formal guarantees-using satisfiability modulo theories (SMT).
\cite{Grune2021} defines \ac{DNN} structure to approximate compositional Lyapunov functions for a system where the approximation error is dependent on the number of neurons in the network. %
\cite{ChenShaoru2021} uses mixed-integer programming in a sampling-based Lyapunov function verification method for piece-wise linear \ac{NN} approximations of nonlinear autonomous systems for estimating their region of attraction. %
\cite{Deka2022} uses Koopman operator theory and sample-based methods to learn a neural network-based Lyapunov function and region of attraction of a given open-loop system.
\cite{Lederer2019} uses \ac{GP}-based approximate dynamic programming with a sample-based method to learn a Lyapunov function and region of attraction of an open loop system. \cite{bonassi2020lstm} analyzes the Input-to-State (ISS) stability of long short-term memory (LSTM) networks by recasting them in the state space form. The authors in \cite{Revay2019, Revay2021} use contraction analysis to design an implicit model structure that allows for a convex parameterization of stable \ac{RNN} models. \cite{Jafarpour2021} uses non-Euclidean contraction theory to establish well-posedness, contraction, and $l_\infty$-Lipschitz properties of implicit \acp{NN}.

\paragraph{Lipschitz properties}\cite{Scaman2018} shows that determining the Lipschitz constant of a \ac{NN} is an NP-hard problem and proposes algorithms to estimate the Lipschitz constant.
Methods for data-driven Lipschitz estimation for controller design and safe policy iteration in ADP are proposed in \cite{chakrabarty2019data,chakrabarty2020safe}.
In \cite{Pauli9319198}, the authors propose a method for training deep feedforward \acp{NN} with bounded Lipschitz constants.
\cite{Fazlyab2022} poses the Lipschitz constant estimation problem for deep neural networks as a semidefinite program (SDP).
\cite{erichson2021lipschitz} shows how to train continuous-time \acp{RNN} with constrained Lipschitz constants to guarantee stability.
\cite{chakrabarty2019robust} develops neural Lipschitz observers with guaranteed performance.

 \paragraph{Robustness and other safety properties}
The authors in~\cite{Wang2016} use the data Jacobian matrix to analyze the behavior of deep neural networks by means of their singular values.
\cite{gehr2018ai2} develops an analyzer for deep convolutional networks with ReLU activations based on an over-approximation that can guarantee  robustness. 
\cite{Jia2020} proposes a method of verification for binarized \acp{NN} using existing boolean satisfiability solvers, which is tested for adversarial robustness to $l_\infty$ perturbations.
\cite{tjeng2019} uses \ac{MILP} formulations to analyze robustness of \acp{NN}. %

\subsubsection{Analysis and verification of learned control policies}

These methods analyze and verify properties of learned control policies in the closed-loop setting.
They are categorized by the considered class of properties as follows.

 \paragraph{Stability}
\cite{ASWANI20131216} provides deterministic stability guarantees for learning-based \ac{MPC} (LBMPC) based on linear state space models.
The authors in~\cite{Hertneck8371312} propose sampling-based probabilistic performance guarantees for approximate \ac{MPC} policies based on the Hoeffding inequality~\cite{Hoeffding1963}.
\cite{Jin2020} uses semidefinite programming to certify stability for model-based \ac{RL} policies.
\cite{Berkenkamp2017} uses Lyapunov stability verification to certify model-based RL policies.
\cite{Berkenkamp2016} uses \ac{GP} to estimate the region of attraction of a closed-loop system.
Recently, \cite{Marchi2022} uses sample-based $\delta$-covers of system domains to ensure input-to-state stability to a set of a closed-loop system when %
an \ac{NN} is used as a state estimator or a closed-loop control law.
\cite{Tsukamoto2021a} develops a neural contraction metric, i.e., an \ac{NN} approximation of a contraction metric, which, coupled with a learning-based controller, ensures exponential convergence to a target trajectory.
\cite{fabiani2021} guarantees stability by combining the worst-case approximation error of the \ac{NN} controller with its Lipschitz constant, utilizing an MILP framework.
\cite{schwan2022} proposes a framework for the stability verification of MILP representable control policies.
\cite{dai2021} simultaneously learns an \ac{NN} controller and Lyapunov function, guided by MILP stability verification, which either verifies stability or gives counter-examples that can help improve the candidate controller and the Lyapunov function.
\cite{zhang2021} learns an \ac{MPC} policy and a ``dual policy,'' which enables them to keep a check on the approximated \ac{MPC}'s optimality online during the control process to filter out suboptimal control inputs and invoke a backup controller with a bounded probability.

 \paragraph{Invariance and other safety properties}
\cite{Zhang2022} uses constrained zonotopes and reachability analysis to determine reachable sets for safety verification of \ac{NN} controllers.
\cite{Robey2020} uses sampled trajectories of a closed-loop system to learn the control barrier function, which can be parameterized by an \ac{NN} and can ensure invariance of the safe set.
\cite{Karg2020} uses an MILP formulation to perform an output range analysis of a trained \ac{NN} controller to guarantee constraint satisfaction and asymptotic stability of the \ac{NN} controller.

\subsubsection{Case study: Verification of a NN controller for a DC-DC power converter}

The following section provides a case study on the verification of a NN controller for a DC-DC power converter. To carry out this verification, we employ the \texttt{EVANQP} framework, developed by \cite{schwan2022}, which utilizes MILPs as its underlying verification technique. By utilizing this framework, we not only demonstrate the closed-loop stability of the NN policy, but we can also validate the satisfaction of constraints in closed-loop operation. The example is adopted from \cite[Section~V]{schwan2022}.

The use of neural network (NN) policies in power converter applications is primarily motivated by the desire to achieve comparable performance to that of an optimal MPC policy while requiring lower computational complexity and memory. Optimal MPC policies are often too computationally demanding to be run in real-time on low-cost microprocessor hardware, making NN policies an attractive alternative. Despite the successful deployment of an NN policy on real hardware by \cite{maddalena2021convertor}, their approach lacks guarantees of stability and constraint satisfaction. In the following, we address these shortcomings.
We will focus on the approach described in \cite{schwan2022} where the baseline policy $\psi_1(\cdot)$ represented by a robust MPC  is approximated by a NN policy $\psi_2(\cdot)$. Specifically, we consider the robust Tube MPC approach proposed by \cite{mayne2005}, which is robust against additive input disturbances in the set $\mathbb{W} = \left\{ w \in \mathbb{R}^m \;\middle|\; \|w\|_\infty \leq \hat{\gamma} \right\}$. By verifying that the NN policy has a worst-case approximation error,
\begin{equation} \label{eq:worst_cast_approximation_error}
    \gamma = \max_{x \in \mathbb{X}} \|\psi_1(x) - \psi_2(x)\|_\infty,
\end{equation}
over a bounded polytopic set $\mathbb{X}$ that is smaller than $\hat{\gamma}$, we can prove that the NN policy $\psi_2(\cdot)$ is asymptotically stable in closed-loop and satisfies constraints on the feasible region of the robust MPC policy $\psi_1(\cdot)$ \cite[Theorem~1]{schwan2022}. Notably, this is achieved by reformulating  \eqref{eq:worst_cast_approximation_error} as a MILP, with the NN policy and the optimal solution map of the robust MPC scheme as MILP constraints.
A wide range of candidate policies including ReLU NNs, optimal solution maps of parametric QPs, and MPC policies can be exactly represented using MILP constraints. We refer to such functions as MILP-representable, as they can be expressed exactly using linear equality and inequality constraints with both continuous and binary decision variables. The MILP-representable functions are piecewise linear in nature, enabling their exact representation using MILP constraints.

\begin{figure}
    \centering
    \begin{subfigure}[b]{0.23\textwidth}
        \centering
        \includegraphics[width=\textwidth]{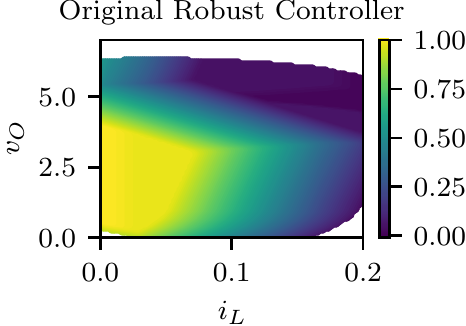}
    \end{subfigure}
    \begin{subfigure}[b]{0.23\textwidth}
        \centering
        \includegraphics[width=\textwidth]{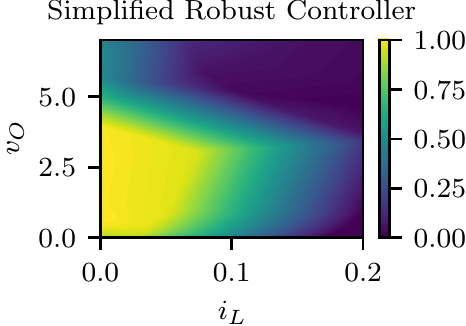}
    \end{subfigure}
    \begin{subfigure}[b]{0.23\textwidth}
        \centering
        \includegraphics[width=\textwidth]{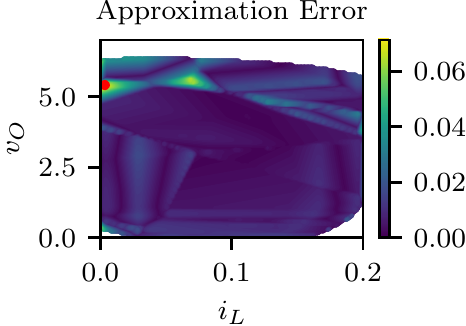}
    \end{subfigure}
    \caption{Original, approximate/simplified robust control policy (top), and approximation error to the original robust MPC control policy (bottom). Taken from \cite{schwan2022}.}
    \label{fig:robust_mpc}
\end{figure}

The model of the DC-DC converter is linearized and discretized, giving us the following two-state $x = (i_L, v_O)$ (current and voltage), and one-input (duty cycle) linear system
\[
    x^+ = Ax + Bu = \begin{bmatrix} 0.971 & -0.010 \\ 1.732 & 0.970 \end{bmatrix}x + \begin{bmatrix} 0.149 \\ 0.181 \end{bmatrix}u.
\]
For the robust MPC we assume the uncertain dynamics
\[
x^+ = Ax + Bu + Bw,
\]
with disturbance set $\mathbb{W} = \left\{ w \in \mathbb{R} \;\middle|\; |w| \leq 0.1\right\}$. We formulate the robust MPC controller
\begin{equation*}
\begin{aligned}
    \min_{\mathbf{z}, \mathbf{v}} \quad & \sum_{i=0}^{N-1}\left(\left\|z_{i}-x_{\mathrm{eq}}\right\|_{Q}^{2}+\left\|v_{i}-u_{\mathrm{eq}}\right\|_{R}^{2}\right)+\left\|z_{N}-x_{\mathrm{eq}}\right\|_{P}^{2} \\
    \text { s.t.} \quad & \forall i=0, \ldots, N-1, \\
    & z_{i+1}=A z_{i}+B v_{i}, \\
    & z_i \in \mathbb{X} \ominus \mathcal{E}, \quad v_i \in \mathbb{U} \ominus K \mathcal{E} \\
    & z_{N} \in \mathbb{X}_{N}, \quad x(0) \in z_{0} \oplus \mathcal{E},
\end{aligned}
\end{equation*}
with steady-state $x_\text{eq}=\begin{bmatrix} 0.05 & 5 \end{bmatrix}^T$, and $u_\text{eq}=0.3379$, state constraints $\mathbb{X} = \left\{ x \in \mathbb{R}^2 \;\middle|\; 0 \leq x_1 \leq 0.2, 0 \leq x_2 \leq 7 \right\}$, input constraints $\mathbb{U} = \left\{ u \in \mathbb{R} \;\middle|\; 0 \leq u \leq 1 \right\}$, terminal set $\mathbb{X}_N$, $\mathcal{E}$ the minimum robust invariant set with respect to a linear feedback gain $K$. The control law is then given by $\psi_1(x)=K(x-z_0^\star(x))+v_0^\star(x)$.
To approximate the robust Tube MPC, we employ a neural network (NN) with 2 hidden layers and 50 neurons each. A saturation layer is added at the end to ensure that the input is clipped between -1 and 1. We use 5000 samples of the Tube MPC uniformly distributed in the feasible region for training the NN, following standard techniques. The NN controller learns an approximation of the MPC policy by minimizing a least-squares loss function. The resulting NN controller and the original Tube MPC can be visualized in Figure~\ref{fig:robust_mpc}.
We employ the \texttt{EVANQP} framework to describe the MPC formulation and NN controller and solve the resulting Mixed-Integer Linear Program (MILP) to determine the worst-case approximation error. Our analysis yields a value of $\gamma=0.073$. As we designed our controller to be robust for input perturbations with a maximum magnitude of $0.1$, our results demonstrate that the NN controller satisfies constraints and converges asymptotically to the steady-state.

\subsection{Learning from physics-informed digital twin simulations}

Advancements in computing and the wide availability of modeling toolkits have yielded high-fidelity simulation software (an essential component of so-called ``digital twins"~\cite{niederer2021scaling}) across a range of domains.
Simulation has become a critical tool for researchers to perform experiments in a scalable, safe, and repeatable manner.
Data generated from digital twins can complement, or offer a powerful alternative to field experiments, which are typically time-consuming, and often require extensive guardrails to ensure expensive equipment/personnel are not subject to harm, or are difficult to repeat.
Some high-fidelity simulators include Modelica (cyber-physical systems~\cite{fritzson2011modelica}, building energy systems~\cite{wetter2014modelica}), CESM (climate models~\cite{kay2015community}), STK (spacecraft~\cite{mccamish2007simulation}), OpenFOAM (fluid flow~\cite{jasak2007openfoam}), and ChainQueen (soft robotics~\cite{hu2019chainqueen}), 
to name a few.

In the prior section on physics-informed system identification, data was assumed to have been obtained from the system under consideration, and an incomplete mathematical representation of the system dynamics is available from domain knowledge (or `physics'). In this subsection, the `physics' is integrated within the data via the data-generating source: the high-fidelity simulator. However, because the simulators comprise software modules that require high modeling complexity and are typically strongly interconnected to one another, directly considering analytical forms of these simulators is unwieldy and impractical. It is generally easier to construct machine learning tools that can directly use data generated by the simulators. %

\subsubsection{Controller design using simulation data}

To ensure that a high-fidelity simulator can be effectively utilized for controller design tasks, it must first be calibrated with real-world data generated by the system of interest. This calibration step is important for ensuring the underlying assumptions and parameter values have been adequately selected to best mimic the true system. However, the complex representation of most high-fidelity simulators implies that the model calibration step can be challenging to solve \cite{paulson2017arbitrary,paulson2019fast}, especially compared to alternative black-box modeling approaches (e.g., deep neural networks) that do not require prior physical knowledge. A natural question arising from this comparison is: Why not directly work with standard machine learning algorithms and forego the high-fidelity simulator altogether? In the context of real-world engineering systems, the answer almost always boils down to a lack of accurate system data. For example, deep learning methodologies have become mainstream solutions in a variety of important applications, such as natural language processing and image classification; however, they often require massive amounts of high-quality labeled training data to surpass human performance. A key advantage of high-fidelity simulators is that they can be calibrated using substantially less training data, which is a direct consequence of the constraints imposed by the underlying first-principles models \cite{raissi2019physics}. Once calibrated, the simulator can then be treated as a \textit{data-generation source} that can be more confidently extrapolated outside of observed training instances. This data generation feature is useful for building control-relevant (or reduced-order) models \cite{ma2016d,paulson2018shaping}, which are critical components of modern model-based control algorithms, such as \ac{MPC}, %
that can handle constrained nonlinear systems with strong multivariate interactions. 

No matter the selected structure, a controller will still depend on design (or tuning) parameters that can strongly affect closed-loop performance and safety.
Historically, these tuning parameters have been selected using a combination of heuristics and/or trial-and-error experimentation on the true system.
To reduce the required testing and validation time, it has been recently proposed to perform controller tuning using  the high-fidelity simulator.  
However, the tuning process is not straightforward due to the computationally expensive nature of high-fidelity simulators.
As such, there has also been significant interest in the development of efficient data-driven automated calibration (or auto-tuning) strategies.
In particular, \ac{BO} has emerged as a powerful approach for handling these types of auto-tuning problems due to its ability to handle expensive black-box functions corrupted by random noise \cite{snoek2012practical,shahriari2015taking,frazier2018tutorial}.
Several recent works have demonstrated the promise of \ac{BO} for auto-tuning of \ac{MPC} \cite{piga2019performance,paulson2020data,sorourifar2021data,lu2020mpc} and other complex control structures \cite{fiducioso2019safe,khosravi2021performance}.
Many interesting extensions of \ac{BO} have also been pursued in the context of auto-tuning, including multi-objective \cite{makrygiorgos2022performance} and robust \cite{paulson2021probabilistically,paulson2022adversarially} formulations.

\subsubsection{Improving generalization of learners via simulations}

A unique opportunity afforded to us by the use of physics-informed simulation tools comes from the ability to \textit{generate useful data} for analysis and design. This generation phase is typically not assumed in most current \ac{PIML} research, where the more standard assumption is that data and domain knowledge pertaining to the `target' system under consideration is available. Contrarily, digital twins comprise parameterized components whose physical parameters or physics-informed structure can be modified in software to generate data from multiple `source' systems that are similar to the target system but not necessarily identical. The implication in modeling and control is that we can use leverage this multi-source dataset to evaluate performance on a range of similar dynamical systems and embed this data into the design pipeline for a target system from which only a few data points are available.
Two classes of \ac{ML} algorithms are naturally suited to learning from multi-source data: (i) meta-learning (also referred to as few-shot learning) and (ii) transfer learning.

\paragraph{Meta-learning}
In meta-learning, two objective functions are typically used in the training phase: an outer-loop loss function for learning commonalities among the multi-source systems and an inner-loop loss function for quickly adapting to a new system with minimal data~\cite{finn2017model}.
At inference, the outer-loop parameters are used to initialize the network, and a few inner-loop iterations are deemed sufficient for adaptation to the target system.
Importantly, the learner structure does not change for the inference task, and the meta-learning algorithm learns to optimize the learning process itself; that is, for a classification problem, the meta-learner may not infer a classification output, but instead may infer a set of hyper-parameters for a classifier network such as loss function parameters or parameters relating to the neural architecture itself~\cite{elsken2020meta}.
\ac{GP} models have been used recently in order to meta-learn predictive models for \ac{MPC}~\cite{arcari2022bayesian} by using dynamic trajectory data from similar systems, and neural networks have been used to meta-learn adaptive control policies in~\cite{richards2021adaptive} for robotics.
Some meta-learning algorithms do not require re-training or bi-level optimization for adaptation.
Instead, they adapt based on contextualization; that is, with the same inputs, the inference changes due to contextual inputs additionally provided to the network.
These context-based meta-learners, such as neural processes and deep kernel networks, have also been investigated recently for parameter learning~\cite{zhan2022calibrating,chakrabarty2022metacontrol} using physics-informed simulators of building energy systems.

\paragraph{Transfer and multi-fidelity learning}
Conversely, transfer learning relies on learning good representations from the multi-source dataset.
At inference, the final network architecture is different from the network that was pre-trained, with the head of the network being inherited from the trained network, while the penultimate layers are altered to fit the learning task on the target system.
Furthermore, the task performed by both the network architectures, e.g., classification, are identical.
So far, the utility of transfer learning in modeling and control has mainly been demonstrated in energy systems~\cite{9147321,xu2020oneformany} where control policies or neural network-based surrogate thermal dynamics models are constructed by using simulations of buildings situated in various climate zones and exhibiting a wide range of architectures. 

A generalized version of the transfer learning problem also arises in controller auto-tuning.
Although auto-tuning strategies applied to the high-fidelity simulator can compensate for the error between the control-relevant model and the simulator, they implicitly assume the error between the simulator and the true system is negligible, which is not always the case in practice.
One way to address this additional source of error is through the framework of \textit{multi-fidelity optimization} wherein we assume access to a family of information sources controlled by a collection of ``fidelity'' parameters that can be generally continuous or discrete.
In the simplest case, we would have one binary fidelity parameter that denotes two distinct but correlated levels, i.e., the simulator and the true system. Due to the general structure of the \ac{BO} framework, several multi-fidelity \ac{BO} methods have been developed in the literature \cite{poloczek2017multi,kandasamy2019multi,song2019general,wu2020practical}, with some being applied directly to the controller auto-tuning problem \cite{sorourifar2021computationally,zhan2022calibrating}.
The main advantage of these types of multi-fidelity optimization methods is their ability to efficiently reconcile discrepancies between the high-fidelity simulator and the true system, which can allow for a significant reduction in the amount of testing required on the true system of interest. 
  
\subsubsection{Sim2Real in Reinforcement Learning} Model-free controllers, e.g. a class of (deep) reinforcement learning algorithms can potentially obtain near-optimal control policies without the need for a mathematical model. This is typically done by the control agent interacting with its environment (or plant) and learning to associate the states to advantageous control actions. The exploration-exploitation trade-off governs this interaction: in order for the agent to gather new knowledge, it needs to explore unknown effects of its actions by choosing them, e.g.,  randomly at times. This feature has three drawbacks for controlling real-world systems. For one, for large state-space systems, the agent may never see all possible states and/or all possible state-action combinations, especially those that may occur infrequently. Second, the amount of data needed for convergence may be large, and consequently the learning times prohibitively long. Third, the agent may choose actions that violate safety  constraints, which may not be permissible in certain plants.

To overcome this issue the sim2real paradigm has been introduced, i.e., the notion that the agent is (pre-)trained in a physically accurate simulator before the deployment on the real system. This way, the agent can be provided with the equivalent of decades of experience and a large variety of potential states. The sim2real paradigm has been successfully explored by developing a variety of simulators for robotics~\cite{robothor, james2019rlbench} and general purpose physics~\cite{ThreeDWorld,Brax}, and demonstrating applications mainly in robotics~\cite{Kadian_2020,sct21iros,truong20navigation,lee20quad} and automotive~\cite{Balaji2020}. Of course, the physics fidelity of the simulators can determine the success and quality of the subsequent deployment, and also its applicability for sim2real. If it is computationally too expensive to run extensive simulation models, they are not suitable to be used in pre-training learning controllers. Here, PIML approaches can be very advantageous in providing physically accurate, yet computationally efficient environments, which can greatly improve the quality of the learning process.

\subsection{Case Study: Violation-aware Bayesian optimization for energy consumption minimization of HVAC with constraints}

\begin{figure}[t]
    \centering
    \includegraphics[width=\columnwidth]{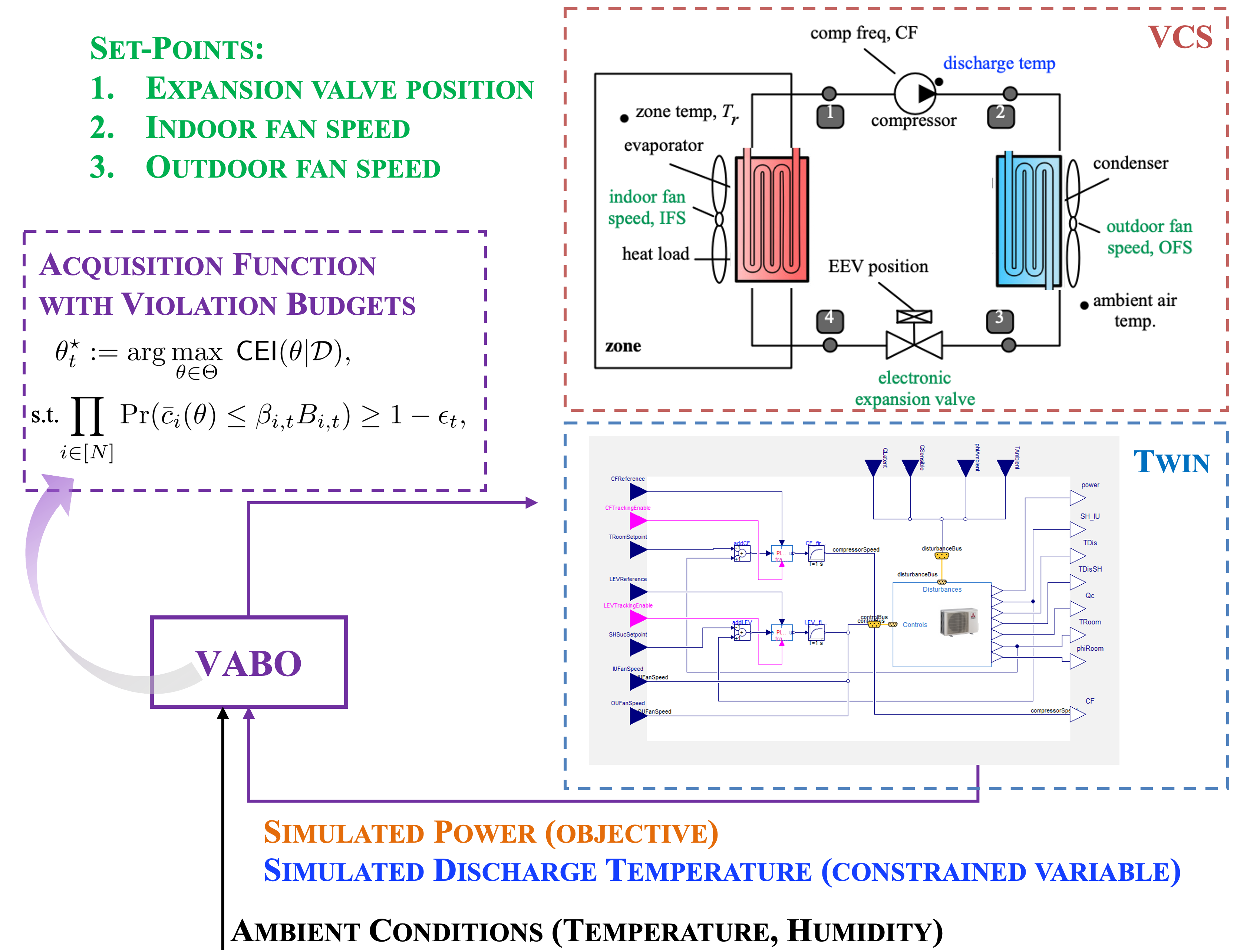}
    \caption{Best feasible solution's power, discharge temperature, cumulative violation cost and the three set-points' evolution. Components of the figure have been taken from~\cite{xu2022vabo} and~\cite{chakrabarty2021accelerating}.}
    \label{fig:sub_plots}
\end{figure}

We consider the problem of safely tuning set-points of an HVAC system as introduced in~\cite{xu2022vabo}, also referred to as a vapor compression system (VCS). A VCS typically consists of a compressor, a condenser, an  expansion valve, and an evaporator. 
Physics-based models of these systems can be formulated as large sets of nonlinear differential-algebraic equations (DAE) to simulate VCS dynamics. To inject realistic refrigerant dynamics and fluid flow, these models contain software blocks that exhibit significant numerical complexity. This motivates directly using  data from VCS digital twin simulations to estimate energy consumption under different operating conditions, to assign set-points to the VCS actuators using data-driven, black-box optimization methods such as Bayesian optimization~\cite{snoek2012practical}.
A high-fidelity digital twin of the VCS was constructed using Modelica~\cite{modelica2017a}; see Fig.~\ref{fig:sub_plots}.  Physics-based models of the compressor, expansion valve, accumulator, and both heat exchangers were interconnected. 
The final DAE model comprises 12114 equations; further modeling details can be found in~\cite{qiao2015a}.

Recklessly changing VCS actuator set-points can drive the system into operating modes that reduce the reliability or lifespan of the system. To avoid these harmful effects, one can add several constraints during the tuning process. One such constraint is the compressor discharge temperature; as high temperatures can result in the breakdown of refrigerant oils and increase wear and tear, shortening the product's lifespan. Furthermore, high temperatures are connected to high pressures, which may cause mechanical fatigue in refrigerant pipes. An advantage is that small constraint violations over a short period of time are acceptable. Authors in~\cite{xu2022vabo} proposed a violation-aware Bayesian optimization (VABO) to minimize energy use in the VCS while trading off constraint violations. 
The feedback loop is closed from compressor frequency to room temperature, leaving the set of three tunable set points  as the expansion valve position and the indoor/outdoor fan speeds. The effects of these set points on power and discharge temperature are not easy to model, and no simple closed-form representation exists. 

Authors in~\cite{xu2022vabo} report that the energy use induced by  VABO decreases slightly faster than generic constrained BO~\cite{gardner2014bayesian} (cBO) and significantly faster than safe BO~\cite{berkenkamp2016safe}. At the same time, the method manages the violation cost well under the violation cost budget.  cBO incurs large discharge temperatures at many iterations because it makes large adjustments to the expansion valve position while maintaining the indoor fan speed at a low value, a combination that is not penalized during exploration. VABO reduces the energy by about $9\%$ compared to the most power-efficient initial safe set-points. Authors also observed that large discharge temperature violations are entirely possible without violation awareness, as demonstrated by  cBO. Safe BO tends to waste a lot of evaluations to enlarge the safe set, which leads to slow convergence; conversely, VABO implicitly encodes the safe set exploration into the acquisition function and only enlarges the safe set when necessary for optimization, while keeping the violation risk small.

\section{Challenges and Opportunities of \ac{PIML} for Control}

This section reflects on open challenges and opportunities of \ac{PIML} methods from various perspectives, including data and prior knowledge requirements, computational demands, safety and performance guarantees, availability of software tools and learning materials, as well as new promising application domains.   
Opportunities of PIML methods in control include:
\begin{enumerate}
    \item Modeling of human behaviors in interactive human-autonomous systems (e.g., autonomous vehicle in mixed traffic). Autonomous system physics is known, but human reaction needs to be learned from real data to model human-in-the-loop systems at scale.
    \item Modeling of high dimensional and distributed physics in multi-physics systems. An example includes combustion processes  where physics is impossible to model compactly by first principles, but there may be good surrogate models that retain core physical properties.
    \item Generation of structured PIML controllers for multi-component systems with awareness of  interconnections of the sub-components. This would lead to improved interpretability and allow for the localization of failures.
    \item Constructing surrogate PIML models for the hard-to-optimize physics-based optimization problems or cost functions. This could significantly speed up the solution of hard optimization problems using cheaper surrogates.
    \item Synthesis of explicit model predictive control policies for large-scale systems leading to a significant reduction in online computational requirements. Thus allowing for the execution of complex control policies in applications with limited communication bandwidths and small sampling rates or enabling deployment on edge devices.
    \item Providing safety and performance guarantees for a broad class of learning-based control methods. Particularly applicable to applications with time-varying dynamics that require online adaptive processes to cope with constant changes in a safe manner. 
    \item Dealing with sim2real gap and allowing for automated tuning of PIML controllers from simulation and experimental data.
    \item Integration of multi-modal inputs into modern control systems. Examples include a combination of video and audio signals with physical measurements processed for downstream decision-making by advanced controls.
\end{enumerate}
Open challenges for PIML methods in control include:
\begin{enumerate}
    \item How to quantify the uncertainty and modeling errors for PIML-based models?  
    \item How to quantify minimal data requirements for training PIML models and controllers?
    \item How to effectively select representative training data for sampling-based PIML approaches?
    \item How to automate the training and hyperparameter tuning process of PIML models?
    \item How to avoid training failures of PIML models getting stuck in local optima. Can we obtain convergence guarantees for certain classes of PIML models?
    \item How to guarantee stability and 
    safety of a real-world system in closed-loop with PIML-based controllers in the presence of noise and plant-model mismatch?
    \item How can verification methods for PIML be scaled up for large-scale or networked systems?
    \item How to reduce the computational requirements of high-fidelity digital twins  without sacrificing accuracy? 
\end{enumerate}

\section{Conclusions}

In the last two decades, the use of machine learning (ML) methods  revolutionized a range of industries, from retail, advertisement, entertainment, healthcare, finance, digital arts, and social networks, to surveillance.
Although highly diverse, these applications have some common denominators, which is that their ML systems  are primarily designed for pattern recognition from multi-modal data sources.
However, as we move towards real-world engineering systems with humans-in-the-loop, such as autonomous vehicles, collaborative robotics,  process control of chemical plants, or power grid, the primary focus is steered toward optimization and control of these dynamical systems with guarantees of safe operation.
In recent years, physics-informed machine learning (PIML) has emerged as a class of methods that systematically combine data-driven ML with physics-based modeling and numerical solvers from engineering.

This tutorial paper provides an overview of the most recent PIML methods applied to the modeling and control of dynamical systems. 
 Specifically, PIML techniques for system identification include structural priors in the architecture of the ML model, matrix factorizations, and physics-informed loss functions.  PIML approaches to control cover 
 learning dynamics models for model predictive control (MPC), learning Lyapunov and barrier functions, differentiable-programming-based control, safe data-driven control, and physics-informed reinforcement learning (RL).
 Obtaining safety and performance guarantees of PIML models and control policies requires rigorous tools verifying properties such as stability, robustness, Lipschitz properties, invariance, and other safety considerations such as constraint satisfaction.  
 The paper also presents methods combining ML with high-fidelity digital twin models suitable for controller design and tuning via meta and transfer learning and dealing with sim2real gap in RL. 
Each major category of PIML methods is accompanied by a representative tutorial case study.

The use of PIML methods in  control presents us with new exciting opportunities that include applications in systems with human-in-the-loop, multi-scale and multi-physics systems, 
and providing benefits such as improved interpretability and modularity, scalability to large-scale complex systems, safety guarantees for adaptive data-driven systems, 
or integration of multi-modal input data in the control systems. 

However, several open questions remain that need to be addressed before the adoption of these methods in real-world applications. These include uncertainty quantification (UQ), data requirements and efficient sampling strategies, automated training, hyperparameter optimization, convergence guarantees, scalability of the verification methods, and computational requirements of high-fidelity physics simulators.

\section*{Acknowledgements}

PNNL authors are supported by the U.S. Department of Energy, through the Energy Efficiency and Renewable Energy, Building Technologies Office under the ``Advancing Market-Ready Building Energy Management by Cost-Effective Differentiable Predictive Control'' project. PNNL is a multi-program national laboratory operated for the U.S. Department of Energy (DOE) by Battelle Memorial Institute under Contract No. DE-AC05-76RL0-1830.
T.X.N. is supported by the National Science Foundation under Grant No.~2138388 and Grant No.~2238296.
EPFL authors are supported by the Swiss National Science Foundation under the NCCR Automation project, grant agreement 51NF40\_180545.

\bibliographystyle{IEEEtran}
\bibliography{references}

\end{document}